\documentclass[english,conference]{IEEEtran}
\usepackage[T1]{fontenc}
\usepackage[latin9]{inputenc}
\usepackage{amsmath}
\usepackage{amssymb}
\usepackage{amsthm,bm}

\makeatletter
\theoremstyle{plain}
\newtheorem{thm}{\protect\theoremname}
\theoremstyle{remark}
\newtheorem{rem}[thm]{\protect\remarkname}

\makeatother

\usepackage{babel}
\usepackage[rgb]{xcolor}
\usepackage{graphics, graphicx}
\usepackage{acronym,balance}

\acrodef{ris}[RIS]{reconfigurable intelligent surface}
\acrodef{aoa}[AOA]{angle of arrival}
\acrodef{aod}[AOD]{angle of departure}
\acrodef{ula}[ULA]{uniform linear array}
\acrodef{fim}[FIM]{Fisher information matrix}
\acrodef{los}[LOS]{line-of-sight}
\acrodef{bs}[BS]{base station}
\acrodef{peb}[PEB]{position error bound}
\acrodef{psd}[PSD]{power spectral density}
\acrodef{ofdm}[OFDM]{orthogonal frequency division multiplexing}

\providecommand{\remarkname}{Remark}
\providecommand{\theoremname}{Theorem}

\begin{document}

\title{Beyond 5G Wireless Localization with Reconfigurable Intelligent Surfaces}

\author{Henk Wymeersch\IEEEauthorrefmark{1} and Beno\^{i}t Denis\IEEEauthorrefmark{2}\\
\IEEEauthorrefmark{1}Department of Electrical Engineering, Chalmers University of Technology, Gothenburg, Sweden\\
\IEEEauthorrefmark{2}CEA-Leti, MINATEC Campus, Grenoble, France\\
e-mail: henkw@chalmers.se, benoit.denis@cea.fr}

\maketitle
\begin{abstract}
5G radio positioning exploits information in both angle and delay, by virtue of increased bandwidth and large antenna arrays. When large arrays are embedded in surfaces, they can passively steer electromagnetic waves in preferred directions of space. Reconfigurable intelligent surfaces (RIS), which are seen as a transformative ``beyond 5G'' technology, can thus control the physical propagation environment. Whereas such RIS have been mainly intended for communication purposes so far, we herein state and analyze a RIS-aided downlink positioning problem from the Fisher Information perspective. Then, based on this analysis, we propose a two-step optimization scheme that selects the best RIS combination to be activated and controls the phases of their constituting elements so as to improve positioning performance. Preliminary simulation results show coverage and accuracy gains in comparison with natural scattering, while pointing out limitations in terms of low signal to noise ratio (SNR) and inter-path interference.
\end{abstract}

\section{Introduction}

Reconfigurable intelligent surfaces (RIS) \acronymused{ris} represent a breakthrough technology
whereby surfaces are endowed with the capability to actively modify
the impinging electromagnetic wave \cite{di_renzo_smart_2019}. RIS
can provide obvious benefits in terms of communication \cite{hu_beyond_2018-1}
and positioning \cite{hu_beyond_2018}, even  though they have been envisaged mostly with specific large continuous surface settings (rather than discrete elements) for the latter application. A recent tutorial is available at \cite{2019Basar}. 
A RIS can operate in three distinct modes: transmission can be achieved by modulating
the phases of the RIS elements \cite{basar_transmission_2019}, reception by providing the RIS with a limited number of RF chains \cite{abdelrahman_taha_deep_2019}, and reflection, which is the most common operating mode  \cite{bjornson_intelligent_2019}, is achieved by real-time control of the RIS elements. 

While an overview of RIS-enabled positioning challenges and opportunities can be found in \cite{wymeersch2019radio}, research has been limited to  RIS operating in the receiver mode \cite{hu_beyond_2018,guidi2019radio}  (with a specific focus on the exploitation of wavefront curvature in the latter) or in reflector mode \cite{he2019positioning}. The work in \cite{he2019positioning} reveals that the RIS can improve position and orientation estimation quality compared to a scatter point, and that phase optimization at the RIS is crucial. Prior to the introduction of RIS, the exploitation of the environment for radio localization has been extensively researched in the multipath localization and mapping literature \cite{LeiJSAC15,leitinger2018belief}. In this literature,  the locations of objects in the environment (surfaces and scatter points) are determined simultaneously with the user's location. Even if these solutions make use of the multipath channel as a constructive source of information as regards to the localization problem geometry, the related electromagnetic interactions (induced by the physical environment) still remain uncontrolled and as such, largely suboptimal from a localization perspective. 

In this paper, we present a Fisher information analysis on a specialized version of \cite{he2019positioning}, with a single transmit and single receive antenna, in order to gain deeper insight into the geometry of the problem. Our analysis reveals that the RIS can be used to control both the direction of Fisher information (thereby essentially improving the geometric dilution of precision) and the amount of information. As another paper contribution inheriting from the latter analysis, we also propose an algorithm that selects the RIS and optimizes their constituting elements, given a (possibly coarse) prior knowledge of the user location, so as to provision the best possible positioning quality and further refine accuracy.

\section{System Model}

\subsection{Geometric Model}
We consider a 2D scenario (see Figure \ref{Fig:Scenario}) with a single antenna transmitter (a \ac{bs}), a single antenna receiver (a user) and a series of $K$ RIS along a wall, each modeled as an $M$-element \ac{ula} with $\lambda/2$ spacing, where $\lambda$ denotes the signal wavelength. The RIS are regularly spaced with inter-RIS spacing $D$. The user and \ac{bs} are assumed to be synchronized\footnote{While this is a strong assumption, it is often used for theoretical analyses as performed here. The synchronization assumption can be removed by considering orthogonal transmissions from multiple \acp{bs}. Alternatively, with multiple surfaces, the user can solve for the unknown position and clock bias \cite{wymeersch20185g}. In either case, 
the analysis still holds.}.
Without loss of generality, we assume the wall is parallel with the x-axis a distance $L$ away from the \ac{bs}. 
The transmitter has known location $[0,0]$, the receiver has unknown location $\mathbf{x}=[x,y]$ and RIS $k$ has an array center at $\mathbf{x}_k=[x_k,L]$. 
Finally, we assume that we can only activate and control up to $\bar{K} \le K$ RIS simultaneously. 

\subsection{Signal and Channel Model}

\begin{figure*}
\centering
\includegraphics[width=0.8\textwidth]{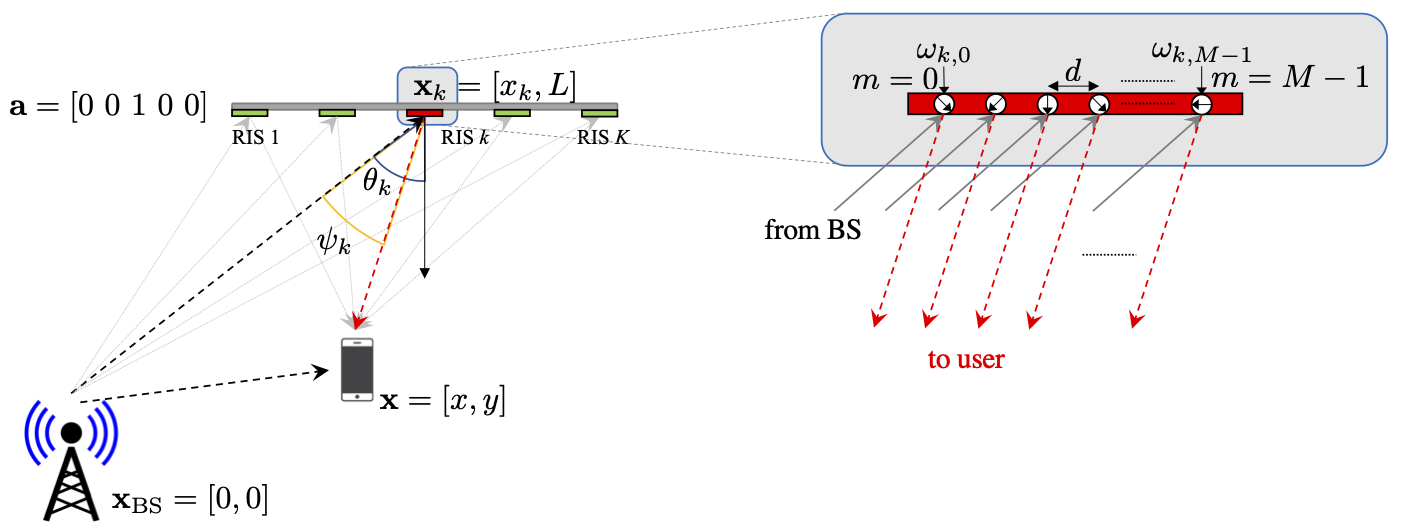}
\caption{RIS-aided positioning scenario with a downlink transmission from a \ac{bs} to a user,  comprising the \ac{los} path and multiple paths via a surface equipped with 1 active/controlled RIS (red) and $K-1$ inactive/uncontrolled RIS (green). {The figure on the right shows the detailed view of the RIS with adjustable phases $\omega_{k,m}$.}} \label{Fig:Scenario}
\end{figure*}

Considering transmission at mm-wave, the received complex baseband signal at the user consists of a \acf{los} signal and a reflected signal \cite{LeiJSAC15}
\begin{align}
    r(t) = \alpha_0 s(t-\tau_0) + \sum_{k=1}^K \alpha_k s(t-\tau_k) + w(t),
\end{align} 
where $s(t)$ is a known \ac{ofdm} signal with average power $P$, $w(t)$ is white Gaussian noise with \ac{psd} $N_0/2$, 
$\tau_0=\Vert \mathbf{x} \Vert / c$, $\tau_k = \Vert \mathbf{x}_k \Vert/c +  \Vert \mathbf{x}_k - \mathbf{x}\Vert/c $, where $c$ denotes the speed of light. The channel gains $\alpha_k$, $k\ge 0$, are modeled geometrically in the mm-wave regime \cite{shahmansoori2017power}: 
\begin{align}
    \alpha_0 & =  e^{-j2\pi f_{c}\tau_{0}}\frac{\lambda}{4 \pi \Vert \mathbf{x} \Vert }\\
\alpha_k & = e^{-j2\pi f_{c}\tau_{k}} \frac{\lambda^2}{16 \pi^2 \Vert \mathbf{x}_k \Vert \Vert \mathbf{x}-\mathbf{x}_k \Vert}\mathbf{h}_k^{\mathsf{T}} \bm{\Omega}_k \mathbf{g}_k,~k>0,
\end{align}
where $f_c$ is the carrier frequency, 
$\mathbf{h}_k$ is the {$M \times 1$} BS-to-RIS response vector with ($m=0,\ldots,M-1$)
\begin{align}
    {h}_{k,m}= \exp(j\pi m \sin(\theta_k))
\end{align}
and $\mathbf{g}_k$ is the {$M \times 1$} LIS-to-UE response vector, with 
\begin{align}
    {g}_{k,m}= \exp(j\pi m \sin(\psi_k)).
\end{align}
%
%
Here, $\theta_k$ denotes the \ac{aoa} of the signal from BS to RIS, and $\psi_k$ denotes the \ac{aod} from RIS to UE, both defined at the RIS with respect to the orthogonal direction (see Fig.~\ref{Fig:Scenario}). 
%
%
%
Finally, the matrix $\bm{\Omega}_k$ is an $M \times M$ diagonal matrix, {which is assumed to be electronically controlled and optimized (e.g., depending on the latest known UE location), of the form} 
\begin{align}
    \bm{\Omega}_{k}= \text{diag}({e^{j \omega_{k,0}}, \ldots, e^{j \omega_{k,M-1}}}). 
\end{align}
We abbreviate $\bm{\omega}=[\bm{\omega}^{\mathsf{T}}_1,\ldots,\bm{\omega}^{\mathsf{T}}_K]^{\mathsf{T}}$, where  $\bm{\omega}_k = \text{diag}(\bm{\Omega}_k)$. 
Our goal is to determine 
$\bm{\omega}$ to provide the best possible positioning quality of the user, based solely on delay measurements (so the dependence of $\alpha_k, k \ge 0$  on $\mathbf{x}$ will not be exploited).

\section{Fisher Information Analysis}
By deriving the \ac{fim} of the unknown parameters, it is possible to derive bounds on the achievable localization accuracy. 
After signal acquisition and conversion to the frequency domain, the observation at the $n$-th subcarrier becomes \cite{Kakkavas19}
\begin{align}
    r[n] = \underbrace{s[n] \sum_{k=0}^{K}\alpha_k e^{-j2 \pi n \tau_k W/(N+1)}}_{f[n]}+ w[n],
\end{align}
where $W$ is the OFDM symbol bandwidth and $s[n]$ is the pilot symbol (with $\mathbb{E}\{ |s[n]|^2\}=E_s = P/W$) on subcarrier $n \in \{ -N/2,\ldots, N/2\}$ {with $N+1$ denoting the total number of subcarriers}. We further assume that the signal spectrum is symmetric.

\subsection{\ac{fim}}
Writing 
the vector of channel gains as $\bm{\alpha}=[\alpha_0,\ldots,\alpha_K]^{\mathsf{T}}$, and the unknowns as $\bm{\eta}=[\mathbf{x}^{\mathsf{T}},\bm{\alpha}^{\mathsf{T}}]^{\mathsf{T}}$, the \ac{fim} is defined as 
\cite{Kay1993}
\begin{align}
 & \mathbf{J}(\bm{\eta})=
 \frac{1}{N_{0}}\sum_{n=-N/2}^{N/2}\Re\left\{ \frac{\partial f[n]}{\partial \bm{\eta}^{\mathsf{H}}} \frac{\partial f[n]}{\partial \bm{\eta}}\right\}.  \label{eq:FIMexpression}
\end{align}
It is readily verified that the entries of the FIM $\mathbf{J}(\mathbf{x},\alpha_l)$ are always equal to zero\footnote{This is because we treat $\alpha_k, k\ge0$ as a separate unknown, independent of $\mathbf{x}$. If the dependence of $\alpha_k$ on $\mathbf{x}$ is considered, $\mathbf{x}$ is the only unknown.}. Hence, we restrict our attention to $\mathbf{x}$ instead of $\bm{\eta}$. We find that 
%
\begin{align}
    \frac{\partial f[n]}{\partial \mathbf{x}} & =  \sum_{k=0}^{K}  \frac{\partial f[n]}{ \partial \tau_k} \frac{\partial \tau_k}{ \partial \mathbf{x}},
    \label{eq:chainRule}
\end{align}
where
\begin{align}
\frac{\partial f[n]}{ \partial \tau_k} & = s[n] \alpha_k e^{-j2 \pi n \tau_k W/(N+1)} \frac{-j2 \pi n  W}{N+1}\\
\frac{\partial \tau_k}{ \partial \mathbf{x}} & = \frac{1}{c} \begin{cases}
    \frac{\mathbf{x}}{ \Vert \mathbf{x}\Vert} & k=0 \\
    \frac{\mathbf{x}-\mathbf{x}_k}{ \Vert \mathbf{x} - \mathbf{x}_k\Vert} & k>0.
    \end{cases}
\end{align}
We now introduce $\mathbf{e}_0=\mathbf{x}/{ \Vert \mathbf{x}\Vert}$ and $\mathbf{e}_k=(\mathbf{x}-\mathbf{x}_k)/{ \Vert \mathbf{x} - \mathbf{x}_k\Vert}$ ($k>0$), which are unit-length vectors pointing from the BS to the user and the $k$-th RIS to the user, respectively. We also introduce 
\begin{align}
    S(\Delta)= \frac{1}{N_0}\sum_{n=-N/2}^{N/2} |s[n]|^2 \left( \frac{2 \pi n  W}{(N+1)c} \right)^2   e^{-j2 \pi n \Delta W/(N+1)}.
\end{align}

With these definitions, substituting \eqref{eq:chainRule} into \eqref{eq:FIMexpression} then yields
\begin{align}
\mathbf{J}(\mathbf{x})=\mathbf{J}^{\text{dir}}(\mathbf{x}) + \mathbf{J}^{\text{int}}(\mathbf{x}),
\end{align}
where $\mathbf{J}^{\text{dir}}(\mathbf{x})$ is the direct path information 
\begin{align}
\mathbf{J}^{\text{dir}}(\mathbf{x})= \sum_{k=0}^{K}|\alpha_k|^2 S(0)\mathbf{e}_k \mathbf{e}^{\mathsf{T}}_k
\end{align}
and $\mathbf{J}^{\text{int}}(\mathbf{x})$ is the inter-path interference information.
\begin{align}
\mathbf{J}^{\text{int}}(\mathbf{x})=  \sum_{k=0}^{K}\sum_{k' \neq k} {\Re\{ \alpha_k  \alpha^*_{k'} S(\tau_{k}-\tau_{k'})\}}
\mathbf{e}_k \mathbf{e}^{\mathsf{T}}_{k'}.
\end{align}
The latter term accounts for the overlapping of multiple paths.  When paths are resolvable, $S(\tau_{k}-\tau_{k'})$ becomes small so that  $\mathbf{J}^{\text{dir}}(\mathbf{x})$ dominates. In that case, each path provides information along the direction $\mathbf{e}_k$ with an intensity $|\alpha_k|^2 \Re\{S(0)\}$. Hence, the performance can be shaped by the RIS through the selection of the directions $\mathbf{e}_k$, as well as the gains $\alpha_k$, $k>0$. When two RIS paths (say $k$ and $k'$) are not resolvable (with $|\tau_k-\tau_{k'}| < 1/W$), then the corresponding gains are added, i.e.,  $\alpha_k+ \alpha_{k'}$, which, due to their approximately random phases leads to an uncontrolled fading-like effect. For that reason, it is not advisable to activate multiple RIS when their paths will not be resolvable in the delay domain. 


\begin{rem}[Data association]
Even when multiple RIS can be found with resolvable path delay differences (i.e., with $|\tau_k-\tau_{k'}| > 1/W$, $\forall k,k'\neq k$), the user must still determine which delay $\tau_k$ corresponds to which RIS with location $\mathbf{x}_k$. Such data association problems are common in multi-path aided positioning and sophisticated tools exist to resolve them \cite{leitinger2018belief}. 
\end{rem}

\subsection{RIS Resource Allocation}

In order to allocate the RIS resources, we consider two variables $\mathbf{a}=[a_1,\ldots,a_K]$ and $\bm{\omega}$, where $a_k \in \{ 0,1 \}$ denotes whether or not  RIS $k$ is activated. When a RIS is not activated (i.e., $a_k=0$), $\bm{\omega}_k=\mathbf{1}_M$, so that the RIS acts as an omnidirectional reflector\footnote{This assumption is likely too pessimistic in terms of inter-path interference, where all the power of the impinging waves would be entirely back-scattered by inactive RIS into all directions indifferently (i.e., including that of the UE), whereas one could reasonably expect more directive effects (e.g., similarly to simple unintentional reflections).}. When a RIS is activated (i.e., $a_k=1$), $\bm{\omega}_k$ should be optimized with respect to the localization performance. As objective function, we consider the \ac{peb} \cite{jourdan2006position}
\begin{align}
\label{eq:PEB}
\mathcal{P}(\mathbf{x}|\mathbf{a},\bm{\omega})=\sqrt{\text{tr}(\mathbf{J}^{-1}(\mathbf{x}))}. 
\end{align}

We can then formulate the following resource allocation problem, in the same spirit as \cite{zhang2016joint}
\begin{subequations}
\label{eq:OPT1}
\begin{align}
\underset{\mathbf{a},\bm{\omega}}{\text{minimize}} & ~~\mathcal{P}(\mathbf{x}|\mathbf{a},\bm{\omega}) \\
\text{s.t.} & ~~ \mathbf{1}^{\mathsf{T}}\mathbf{a} \le \bar{K}\\
& ~~d_{\text{min}}(\mathbf{a}) > {c}/{(WD)},
\end{align}
\end{subequations}
where we recall that $\bar{K}$ is the maximum number of RIS that can be activated simultaneously and $D$ is the inter-RIS spacing. The function $d_{\text{min}}(\mathbf{a}): \{ 0,1\}^{K} \to \mathbb{N}$ returns the minimum index-wise distance between two consecutive ones in $\mathbf{a}$ (i.e., $d_{\text{min}}([ 0~ 1~ 1])=1$ and $d_{\text{min}}([ 1 ~0~1])=2$), so as to prevent from multi-path overlap.

Given $\mathbf{a}$, the optimization over $\bm{\omega}$ is straightforward.
When $a_k=1$, we should then set $\bm{\omega}_k$ to maximize $|\alpha_k|^2$, with 
\begin{align}
    |\alpha_k|^2 & \propto |\mathbf{h}_k^{\mathsf{T}} \bm{\Omega}_k \mathbf{g}_k |^2 \\
    & = \left| \sum_{m=0}^{M-1}e^{j\omega_{k,m}}e^{j\pi m \sin(\theta_k)} e^{j\pi m \sin(\psi_k)} \right|^2,
\end{align}
which is maximized for 
\begin{align}
    \omega_{k,m}=-\pi m (\sin(\theta_k)+\sin(\psi_k)),
\end{align}
or, equivalently, $\bm{\omega}_k=(\mathbf{g} \odot \mathbf{h})^*$, where $\odot$ denotes the Hadamard product. 
Under this choice, 
\begin{align}
    |\alpha_k|^2 = \frac{\lambda^4 M^2}{16^2 \pi^4 \Vert \mathbf{x}_k \Vert^2 \Vert \mathbf{x}-\mathbf{x}_k \Vert^2}.
\end{align}
In other words, we observe an SNR gain of $M^2$ by appropriate choice of the weights $\bm{\omega}_k$. 

The optimization problem \eqref{eq:OPT1} then becomes 
\begin{subequations}
\label{eq:OPT2}
\begin{align}
\underset{\mathbf{a}}{\text{minimize}} & ~~\mathcal{P}(\mathbf{x}|\mathbf{a},\bm{\omega}(\mathbf{a}))  \\
\text{s.t.} & ~~ \mathbf{1}^{\mathsf{T}}\mathbf{a} \le \bar{K}
\\
& ~~d_{\text{min}}(\mathbf{a}) > {c}/{(WD)},
\end{align}
\end{subequations}
which is combinatorial in nature. For small $K$, optimizing over $\mathbf{a}$ can be achieved through exhaustive search. 

Intuitively, the RIS selection should be used to improve the rank of $\mathbf{J}(\mathbf{x})$.  Given a first-step RIS activation uniquely based on geometric considerations (i.e., the rank of $\mathbf{J}(\mathbf{x})$), the choice of $\omega_{k,m}^{*}$ thus naturally results in concentrating the reflected signal energy towards the UE. 

\begin{rem}[User location uncertainty] The resource allocation was performed for a given user location $\mathbf{x}$, which is exactly the quantity we want to estimate \emph{in fine}. To avoid this chicken-and-egg problem, we can substitute in \eqref{eq:OPT2} an estimate\footnote{The uncertainty affecting prior UE location information could be taken into account while optimizing the RIS elements, for instance by computing a Bayesian version of the positioning error bound  or simply by adding optimization constraint so that the reflected power is still concentrated towards the user, but with a spatial margin depending on this uncertainty, thus controlling the probability of pseudo-beam misalignment (similar to \cite{Koirala_Throughput_2018} for mm-wave joint localization and communication services). {Assessing the sensitivity of the proposed approach to prior location uncertainty falls out of the scope of this paper and is left for future work.}} of $\mathbf{x}$, or  by replacing $\mathcal{P}(\mathbf{x}|\mathbf{a},\bm{\omega}(\mathbf{a}))$ with $\max_{\mathbf{x} \in \mathcal{X}} \mathcal{P}(\mathbf{x}|\mathbf{a},\bm{\omega}(\mathbf{a})) $ (when we know a priori a region $\mathcal{X}$ of $\mathbf{x}$) or with  $\mathbb{E}_{\mathbf{x}}\{ \mathcal{P}(\mathbf{x}|\mathbf{a},\bm{\omega}(\mathbf{a})) \}$ (when we know an a priori density $p(\mathbf{x})$)  \cite{shahmansoori2017power,dai2014distributed}. 
\end{rem}

\section{Comparison with passive objects}
It is instructive to compare the performance of the RIS with standard passive objects with known location, such as a reflecting surface and a scatter point. 

\subsection{Reflecting Surface}
\subsubsection{Model} \label{sec:ModelSurface}
We consider a reflecting surface extending from $[h_1,L]$ to $[h_2,L]$ (see also Figure \ref{Fig:Scenario}), both known to the user. With this surface, we can associate a virtual anchor at $\mathbf{x}_{\text{VA}}=[0,2L]$, corresponding to the location of the BS, reflected with respect to the surface. Then, given the user location $\mathbf{x}$, we can associate an incidence point $\mathbf{s}(\mathbf{x})$ on the surface, which is given by the intersection of the surface and the line between $\mathbf{x}_{\text{VA}}$ and $\mathbf{x}$. We denote by $I\{\mathbf{x}\}\in \{0,1\}$ whether or not $\mathbf{s}(\mathbf{x})$ exists: when the finite surface and the line between $\mathbf{x}_{\text{VA}}$ and $\mathbf{x}$ have no intersection point, then $I\{\mathbf{x}\}=0$.

The received signal becomes 
\begin{align}
    r(t) = \alpha_0 s(t-\tau_0) +  \alpha_r s(t-\tau_r) + w(t), \label{eq:modelSurface}
\end{align}
where now $\tau_r= \Vert \mathbf{s}(\mathbf{x}) \Vert/c +  \Vert \mathbf{s}(\mathbf{x}) - \mathbf{x}\Vert/c = \Vert \mathbf{x}_{\text{VA}} - \mathbf{x}\Vert/c$ is the arrival time of the reflected path and 
\begin{align}
    \alpha_r =I\{\mathbf{x}\} \frac{\lambda\Gamma}{4\pi\Vert \mathbf{x}_{\text{VA}} - \mathbf{x}\Vert},
\end{align}
in which $0 \le \Gamma \le 1$ is the reflection coefficient. 

\subsubsection{FIM}
Under the assumption that the two paths in \eqref{eq:modelSurface} can be resolved, then 
\begin{align}
 \mathbf{J}(\mathbf{x}) 
& \approx S(0)|\alpha_0|^2 \frac{\mathbf{x}}{\Vert\mathbf{x}\Vert}\frac{\mathbf{x}^{\mathsf{T}}}{\Vert\mathbf{x}\Vert}\\
& + S(0)|\alpha_r|^2 \frac{\mathbf{x}-\mathbf{x}_{\text{VA}}}{\Vert\mathbf{x}-\mathbf{x}_{\text{VA}}\Vert}\frac{(\mathbf{x}-\mathbf{x}_{\text{VA}})^{\mathsf{T}}}{\Vert \mathbf{x}-\mathbf{x}_{\text{VA}}\Vert}. \nonumber
\end{align}
Comparing to the RIS case, we note that generally $|\alpha_r|^2 \gg |\alpha_k|^2$ (since, loosely speaking, $|\alpha_k|^2$ decays with the 4-th power of the distance, while $|\alpha_r|^2$ only decays with the 2-nd power of the distance), so that the reflecting surface can provide higher FIM intensity. On the other hand,  when $I\{\mathbf{x}\}=0$, the reflected path is not present, and the FIM becomes degenerate since we only receive information in one direction. In terms of estimation, this corresponds to a non-resolvable positioning problem.

\subsection{Scatter Point}
\subsubsection{Model}
We consider a scatter point located at $\mathbf{s}=[s,L]$, known to the user. Note that, different from  Section  \ref{sec:ModelSurface}, $\mathbf{s}$ is not a function  of $\mathbf{x}$.  The received signal becomes 
\begin{align}
    r(t) = \alpha_0 s(t-\tau_0) +  \alpha_s s(t-\tau_s) + w(t), \label{eq:ModelSP}
\end{align}
where now $\tau_s= \Vert \mathbf{s} \Vert/c +  \Vert \mathbf{s} - \mathbf{x}\Vert/c $ is the arrival time of the scattered path and 
\begin{align}
    \alpha_s = \frac{\lambda\sqrt{\sigma}}{(4\pi)^{3/2}\Vert \mathbf{s} \Vert\Vert \mathbf{s} - \mathbf{x}\Vert},
\end{align}
in which $\sigma \ge 0$ is the object radar cross section (expressed in $\text{m}^2$).

\subsubsection{FIM}
Under the assumption that the two paths in \eqref{eq:ModelSP} can be resolved, then 
\begin{align}
 \mathbf{J}(\mathbf{x}) 
& \approx S(0)|\alpha_0|^2 \frac{\mathbf{x}}{\Vert\mathbf{x}\Vert}\frac{\mathbf{x}^{\mathsf{T}}}{\Vert\mathbf{x}\Vert}\\
& + S(0)|\alpha_s|^2 \frac{\mathbf{x}-\mathbf{s}}{\Vert\mathbf{x}-\mathbf{s}\Vert}\frac{(\mathbf{x}-\mathbf{s})^{\mathsf{T}}}{\Vert \mathbf{x}-\mathbf{s}\Vert}.
\end{align}
In contrast to the reflecting surface, the scatter point can ensure a full-rank FIM for all $\mathbf{x}$. On the other hand, $|\alpha_s|^2$ may be small, when either $\Vert \mathbf{s} \Vert$ or $ \Vert \mathbf{s} - \mathbf{x}\Vert$  are large. In contrast to the RIS, a scatter point can control neither the direction of information, nor the intensity of information. 

\section{Numerical Results}

\begin{figure}[t!]
\centering
\includegraphics[width=1\columnwidth]{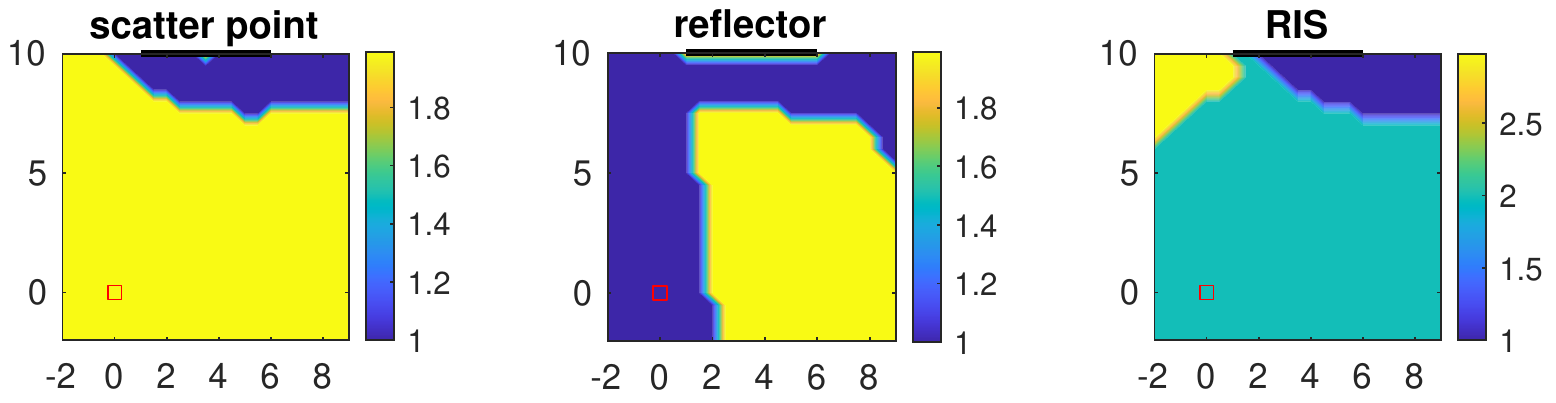}
\caption{Number of resolvable paths as a function of the user location for a scatter point, a surface, and the optimized RIS with $\bar{K}=1$. $W=100$ MHz.} \label{Fig:paths}
\end{figure}

\begin{figure}[t!]
\centering
\includegraphics[width=1\columnwidth]{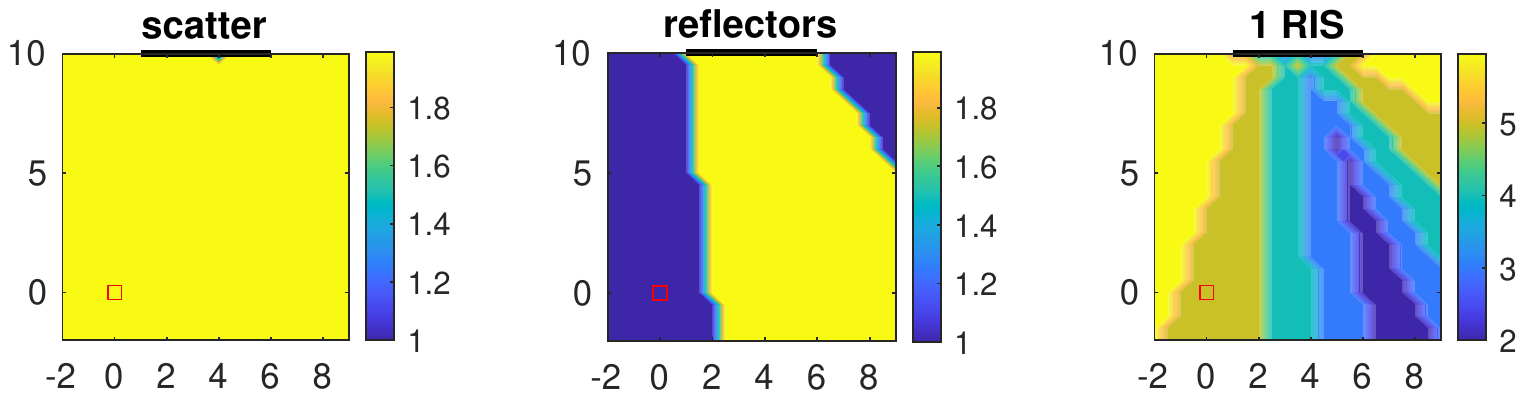}
\caption{Number of resolvable paths as a function of the user location for a scatter point, a surface, and the optimized RIS with $\bar{K}=1$. $W=1$ GHz.} \label{Fig:paths1000}
\end{figure}
\subsection{Simulation Scenario}
We consider an OFDM system at $f_c= 28~\text{GHz}$ with $W \in \{ 100~\text{MHz}, 1~\text{GHz}\}$ total bandwidth, using 129 subcarriers, and QPSK pilots.  This leads to  a distance resolution of 3 m (resp.~30 cm) and a maximum unambiguous range of approx.~387 m (resp.~38.7 m). 
The transmit power is set to 1 mW. We consider up to 5 RIS, each with $M=100$ elements, extending from $[1,10]$ to $[6,10]$ with an inter-RIS spacing of $D= 1 ~\text{m}$. For comparison, we also show results for a reflecting surface extending with $h_1=1$, $h_2=6$, with reflection coefficient $\Gamma=0.3$. Finally, we also compare with a scatter point located at $\mathbf{s}=[3.5, 10]$ with RCS $\sigma = 0.01~\text{m}^2$. We ignore near-field effects. 

\subsection{Discussion}

\subsubsection{Number of Resolvable Paths} In Fig.~\ref{Fig:paths}--\ref{Fig:paths1000}, we show the number of resolvable paths for $W=100~\text{MHz}$ and $W=1~\text{GHz}$. Close to the scatterer, the paths are not resolvable for the smaller bandwidth. The same is true close to the reflector. In addition, the reflector only provides a NLOS path in a certain region (where $I\{\mathbf{x}\}=1$). In the case of the RIS, more paths can become available (up to 3 for 100 MHz and up to 6 for 1 GHz).

\subsubsection{Spatial PEB} 
\begin{figure}
\centering
\includegraphics[width=1\columnwidth]{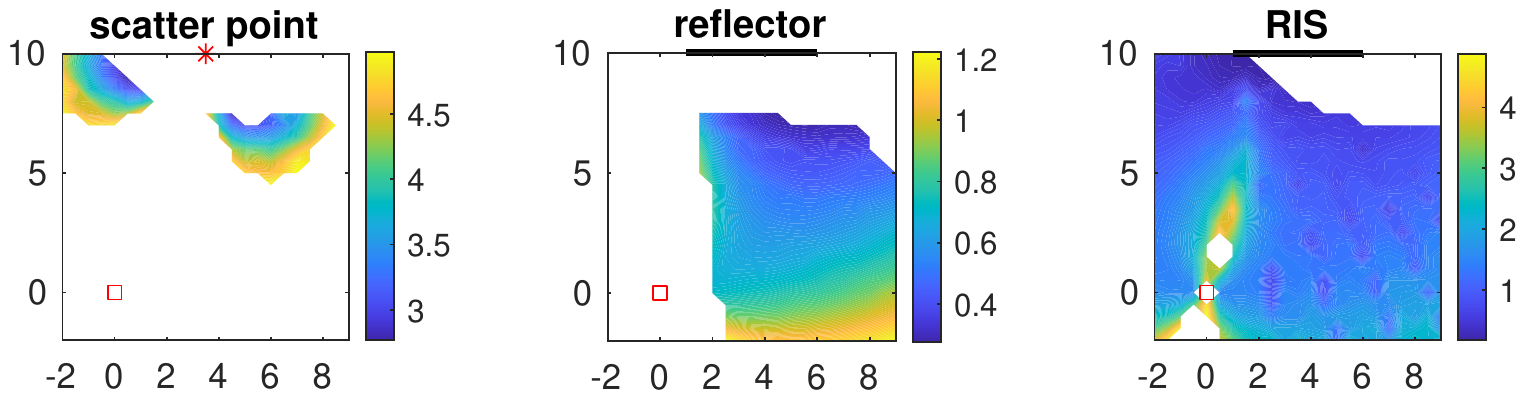}
\caption{PEB as a function of the user location for  a scatter point, a surface, and the optimized RIS with $\bar{K}=1$. $W=100$ MHz.} \label{Fig:mycontours}
\end{figure}

\begin{figure}
\centering
\includegraphics[width=1\columnwidth]{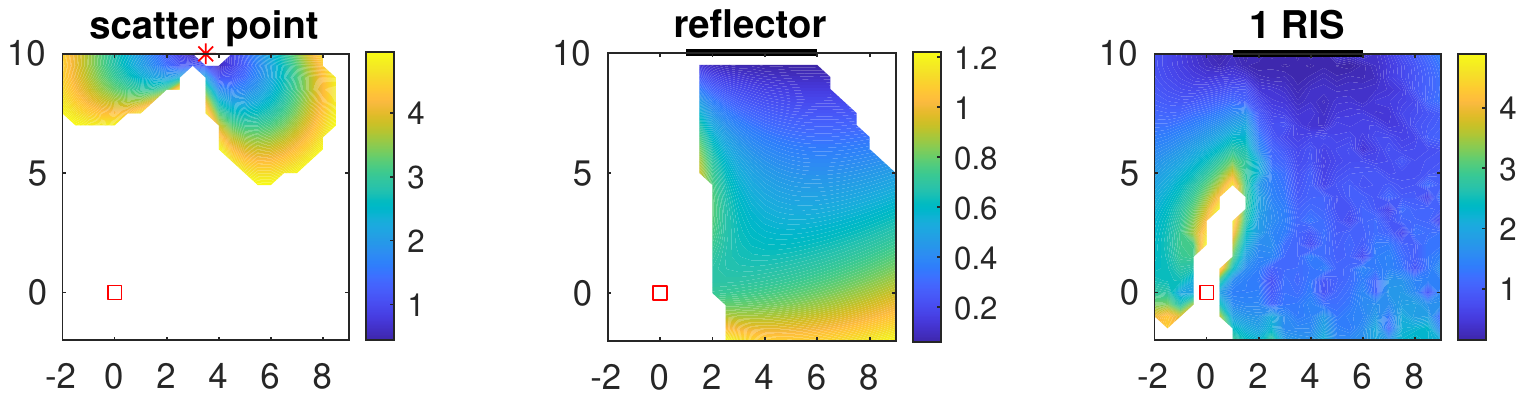}
\caption{PEB as a function of the user location for a scatter point, a surface, and the optimized RIS with $\bar{K}=1$. $W=1$ GHz.} \label{Fig:mycontours1000}
\end{figure}

Fig.~\ref{Fig:mycontours}--\ref{Fig:mycontours1000} show the spatial distribution of the PEB (capped at 5 m) for the scatter point, reflector and a single RIS ($\bar{K}=1$). For the scatter point, the PEB takes on low values close to the scatter point, but not so close that the paths are no longer resolvable. For the reflector, the PEB is low when the reflected path is available. For the optimized RIS, a PEB less than 5 m can be achieved throughout most of the deployment region. When the bandwidth is larger, all scenarios benefit from better delay resolution, in particular close to the object. %
{For all cases, we note that there are white regions where} {the PEB is worse} {than 5 m. We discern 2 cases: (i) where one can not issue a unique positioning result (when only 1 path can be resolved), so that  $\text{PEB}=+\infty$, (ii) where the PEB exceeds the 5 m threshold, due to a poor diversity of information directions and/or poor SNR (despite a sufficient number of resolved paths).}

\subsubsection{PEB CDF} 
\begin{figure}
\centering
\includegraphics[width=1\columnwidth]{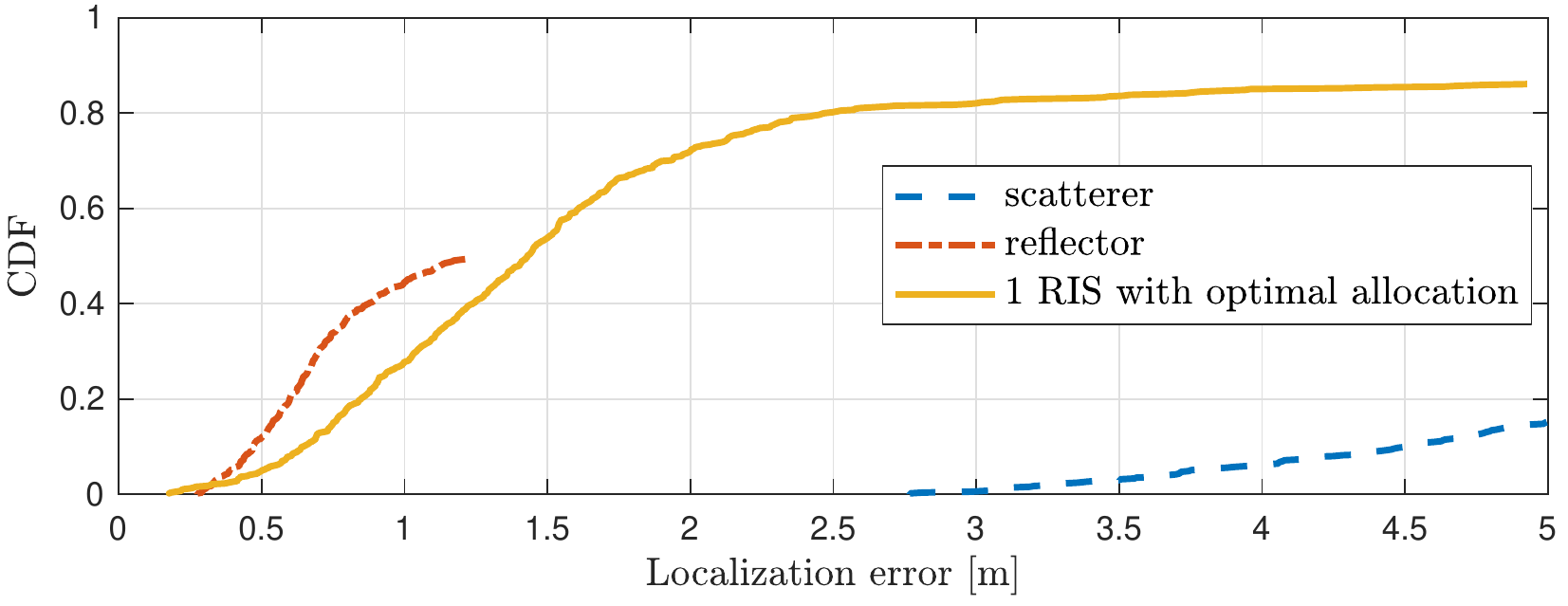}
\caption{CDF of the PEB within the deployment region for $\bar{K}\ = 1$. $W=100$ MHz.} \label{Fig:CDF}
\end{figure}

\begin{figure}
\centering
\includegraphics[width=1\columnwidth]{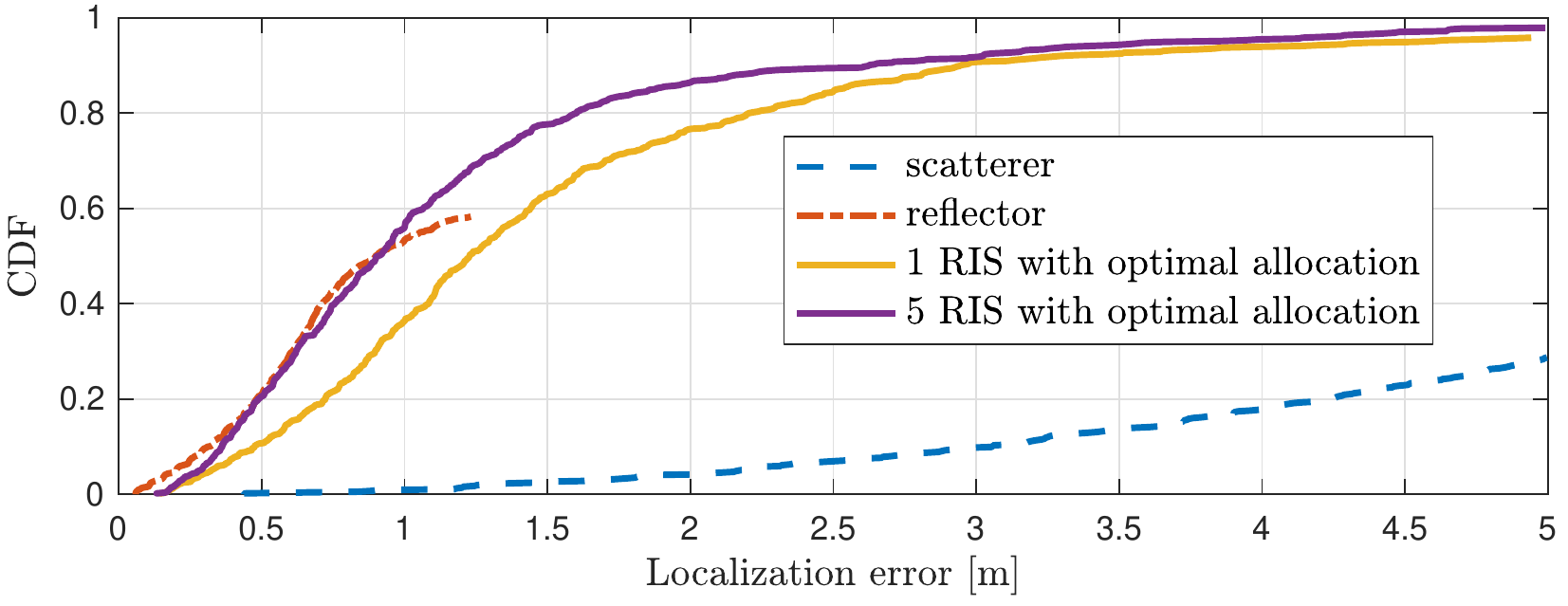}
\caption{CDF of the PEB within the deployment region for $\bar{K}\ \in \{1,5\}$. $W=1$ GHz.} \label{Fig:CDF1000}
\end{figure}

In Fig.~\ref{Fig:CDF}--\ref{Fig:CDF1000} we visualize the PEB results as a cumulative density function (CDF). For 100 MHz bandwidth, the reflector offers sub-meter performance for about 45\% of the deployment region. The RIS covers over 80 \% with PEB less than 2.5 meters. Increasing the bandwidth to 1 GHz  only slightly improves the performance (since the transmit power is the same and the OFDM symbols are shorter). Now $\bar{K}=5$ can be evaluated, since the RIS paths can be resolved. We notice a significant performance improvement, similar to the reflecting surface, but with larger coverage. 

\begin{figure}[t!]
\centering
\includegraphics[width=1\columnwidth]{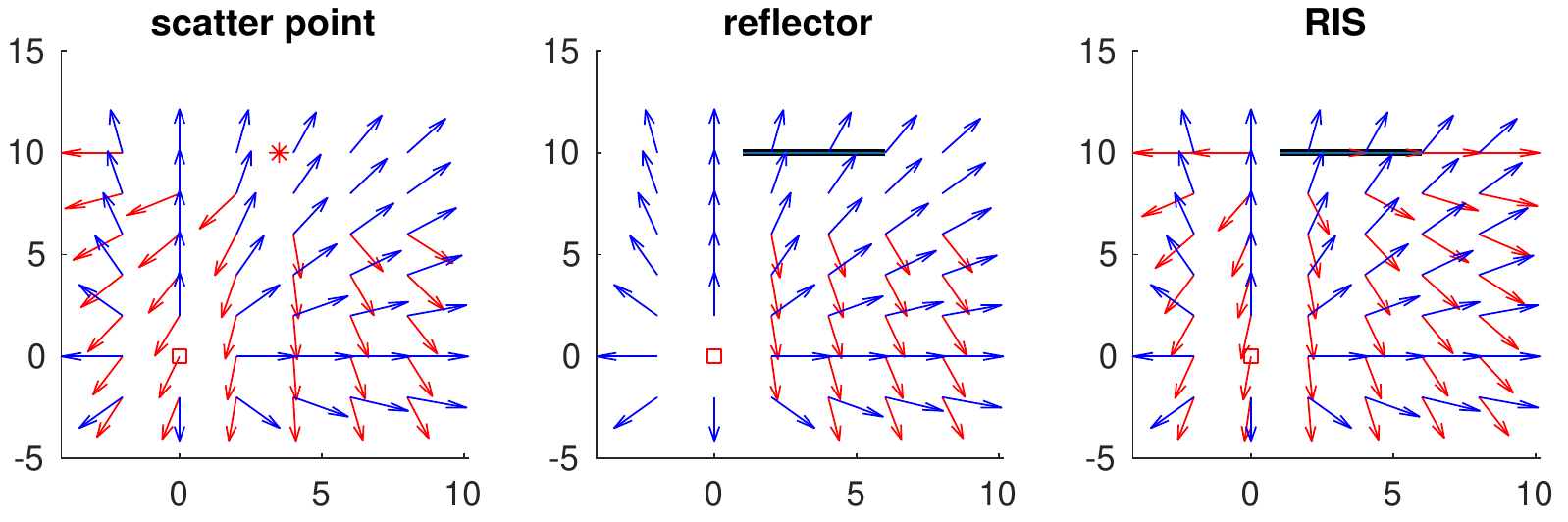}
\caption{Directions of information in the FIM for $W=100$ MHz. {The arrows show (in blue) the direction  $\mathbf{x}/\Vert \mathbf{x} \Vert$ from the BS and (in red), when the} {secondary path is present, the directions $({\mathbf{x}-\mathbf{s}})/{\Vert\mathbf{x}-\mathbf{s}\Vert}$ (left figure), } {$({\mathbf{x}-\mathbf{x}_{\text{VA}}})/{\Vert\mathbf{x}-\mathbf{x}_{\text{VA}}\Vert}$  (middle figure) and $(\mathbf{x}-\mathbf{x}_k)/{ \Vert \mathbf{x} - \mathbf{x}_k\Vert}$ (right figure). } } \label{Fig:direction}
\end{figure}

\subsubsection{Direction of information} Finally, we visualize the directions of information in the FIM for the three cases in Fig.~\ref{Fig:direction}. For the scatter point and the reflecting surface, these directions are deterministic, while for the RIS, they can be optimized by choice of the selection vector $\mathbf{a}$.


\section{Conclusions}

Beyond single-BS multipath-aided positioning capabilities, RISs provide a unique opportunity to control the propagation channel and hence, to guarantee a theoretical positioning accuracy level regardless of the occupied UE location (i.e., achieving near-homogeneous localization quality-of-service over space in a given scene). In this paper, we have analyzed the RIS-aided positioning problem from a FIM perspective, before introducing a simple solution to activate and optimize the best RISs accordingly. Significant theoretical performance gains have thus been illustrated when activating one single RIS already in terms of both coverage (resp.~PEB)  when compared to a single passive reflector (resp.~a single passive scatterer). Obviously, the RIS-based approach is naturally hampered by lower SNR, unless the number of elements in the RIS is large. Finally, we have also adversely observed limitations at large numbers of activated RISs, due to unresolved multipath components. 

Although far-field approximations have been made throughout the paper, the far-field distance exceeds the size of the deployment area in the considered region. Hence future work should explore near-field phenomena and more specifically, the possibility to exploit the curvature of wavefronts impinging onto (resp.~departing from) array-based RIS (e.g., with uniform linear or rectangular arrays), similar to \cite{guidi2019radio}, but in reflection mode. Beyond, we aim to extend our FIM analysis (and the resulting FIM-based RIS selection/optimization problem) to the estimation of various location-dependent metrics, such as received signal strength (RSS), AOD, AOA, for which the geometric dilution of precision can still be controlled. The uncertainty of UE location should also be taken into account to develop suitable procedures that aim at jointly refining UE position and RIS configuration.

\section*{Acknowledgment}
This work was supported, in part, by the Swedish Research Council (VR) under project No.~2018-03701. 

\balance
\bibliographystyle{ieeetr}
\bibliography{references}

\end{document}